\documentclass[a4paper,10pt]{article}
\usepackage[centertags]{amsmath}
\usepackage{amsmath}
\usepackage[dutch,british]{babel}
\usepackage{amsfonts}
\usepackage{amssymb}
\usepackage{amsthm}
\usepackage{newlfont}
\usepackage{color}

 \usepackage{bbm}

\usepackage{stmaryrd}
\usepackage{mathrsfs}
\usepackage{pstricks,pst-node}
\usepackage{palatino}  
\usepackage{fancyhdr}
\usepackage{graphicx}
\usepackage{fancybox}
\usepackage{multibox}
\usepackage{geometry}

\theoremstyle{plain}
  \newtheorem{theorem}{Theorem}[section]
  
  \newtheorem{proposition}[theorem]{Proposition}
  \newtheorem{lemma}[theorem]{Lemma}
  
    \newtheorem{assumption}[theorem]{Assumption}
\theoremstyle{definition}

\theoremstyle{remark}

\numberwithin{equation}{section}

\DeclareMathOperator{\Tr}{Tr}

\newcommand\otimesal{\mathop{\hbox{\raise 1.6 ex
  \hbox{$\scriptscriptstyle\mathrm{al}$}
\kern -0.92 em \hbox{$\otimes$}}}}
\newcommand\oplusal{\mathop{\hbox{\raise 1.6 ex
  \hbox{$\scriptscriptstyle\mathrm{al}$}
\kern -0.92 em \hbox{$\oplus$}}}}
\newcommand\Gammal{\hbox{\raise 1.7 ex
\hbox{$\scriptscriptstyle\mathrm{al}$}\kern -0.50 em $\Gamma$}}
\renewcommand\i{\mathrm{i}}

\let\al=\alpha   \let\ep=\epsilon

  \let\ga=\gamma 
 \let\la=\lambda  
\let\si=\sigma

   \let\Om=\Omega

\newcommand{\caB}{{\mathcal B}}
\newcommand{\caC}{{\mathcal C}}

\newcommand{\caH}{{\mathcal H}}

\newcommand{\caP}{{\mathcal P}}

\newcommand{\caU}{{\mathcal U}}
\newcommand{\caV}{{\mathcal V}}

\newcommand{\bbC}{{\mathbb C}}

\newcommand{\bbE}{{\mathbb E}}

\newcommand{\bbN}{{\mathbb N}}

\newcommand{\bbR}{{\mathbb R}}

\newcommand{\bbT}{{\mathbb T}}

\newcommand{\bbZ}{{\mathbb Z}}

\newcommand{\opunit}{\text{1}\kern-0.22em\text{l}}

\newcommand{\Hi}{{\mathcal{H}}}

\newcommand{\e}{{\mathrm e}}

\renewcommand{\d}{{\mathrm d}}

\newcommand{\Dom}{\mathrm{Dom}}

\newcommand{\beq}{ \begin{equation} }
\newcommand{\eeq}{ \end{equation} }
\newcommand{\bet}{ \begin{theorem} }
\newcommand{\eet}{ \end{theorem} }

\newcommand{\baq}{\begin{eqnarray}}
\newcommand{\eaq}{\end{eqnarray}}

 \newcounter{smallarabics}
\newenvironment{arabicenumerate}
{\begin{list}{{\normalfont\textrm{\arabic{smallarabics})}}}
  {\usecounter{smallarabics}\setlength{\itemindent}{0cm}
  \setlength{\leftmargin}{5ex}\setlength{\labelwidth}{4ex}
  \setlength{\topsep}{0.75\parsep}\setlength{\partopsep}{0ex}
   \setlength{\itemsep}{0ex}}}
{\end{list}}

\newcounter{smallroman}

\newcommand{\ben}{\begin{arabicenumerate}}
\newcommand{\een}{\end{arabicenumerate}}

\newcommand{\norm}{ \|}
\newcommand{\str}{ |}

\newcommand{\tor}{ \bbT^d }
\newcommand{\lat}{ \bbZ^d }

\begin{document}
\begin{center}
{\Large \bf{Diffusive behavior from
 a quantum master equation}} \\
\vspace{10pt}

{\bf   Jeremy Clark\footnote{
email: {\tt  jclark@mappi.helsinki.fi    }} }\\
\vspace{10pt} 
{\it   Department of Mathematics \\
 University of Helsinki \\ 
P.O. Box 68, FIN-00014,  Finland 
} \\

\vspace{15pt}
{\bf   W.  De Roeck\footnote{
email: {\tt
 w.deroeck@thphys.uni-heidelberg.de}}  }\\
\vspace{10pt} 
{\it   Institut f\"ur Theoretische Physik  \\ Universit\"at Heidelberg \\
Philosophenweg 19,  \\
D69120 Heidelberg,  Germany 
} \\

\vspace{15pt}

{\bf   Christian Maes \footnote{
email: {\tt christian.maes@fys.kuleuven.be }}  }\\
\vspace{10pt} 
{\it  Instituut voor Theoretische Fysica \\
K.U.Leuven \\ 
Celestijnenlaan 200D, Belgium
} \\


\end{center}

\begin{abstract}
We study a general class of translation invariant quantum
Markov evolutions for a particle on $\bbZ^d$.  The evolution consists of free flow, interrupted by scattering events. 
We assume
spatial locality of the scattering events and exponentially fast relaxation of the momentum distribution. It is shown that
the particle position diffuses in the long time limit.  This generalizes standard results about central limit theorems for classical (non-quantum) Markov processes. 

\end{abstract}

 \large{
\section{Introduction}
A classical problem in the study of dynamical systems is understanding diffusive behavior for some properly rescaled variable.
When starting from a Hamiltonian dynamics, that often proceeds in
two steps. First there is a the identification of relevant space-time scales under which certain variables obey a reduced autonomous description.  That specifies the limit starting from a microscopic dynamics and leading to a translation invariant master or Boltzmann-type equation, e.g. as the result of a weak coupling
or a low density approximation, \cite{spohn,DZ,pul}. Already there
some irreversible behavior may be exhibited.
Additionally, a second step can further specify the
irreversible properties of a more restricted set of degrees of
freedom.  

The present paper deals with the second step in a quantum set-up, taking
for granted a form of the master equation for the reduced
description of a quantum particle hopping on the lattice. 
We imagine a translation invariant law of motion wherein the free
Hamiltonian evolution is interrupted by scattering events from
interactions with the environment. The effective or resulting
description is that of a Markovian open system. In quantum
mechanics, Lindblad equations take the place of Langevin or
Fokker-Planck equations in classical probability theory describing
a dynamical system under the influence of an idealized noisy
environment, cf. \cite{AF}. This Lindblad equation is  a master
equation for the evolution of the density matrix. The models we
study in the present paper are translation invariant Markovian
evolutions for a quantum particle on $\bbZ^d$.  Its state is
described via a density matrix $\rho_t$ for which, in position
representation, the diagonal $\rho_t(x,x)$ gives the probability of the spatial location being $x$ at time $t$. 
The
specific derivation of such master equations, very much like a
linear Boltzmann equation, starting from the unitary evolution of
a particle in contact with a reservoir is not the subject of the
paper.  Recently, some rigorous derivations of this type have been carried out successfully by several authors, e.g.\  \cite{erdosyau, erdoseng, DFP,DF}.  We do not discuss these derivations here, and we just restrict ourselves to the remark that not all of them lead to a quantum Markov process in the sense meant in this paper, namely as a Markovian equation for the density matrix of the particle, also called Lindblad equation.  This comes about because  the object that admits a scaling limit is often a Wigner function rather than a density matrix (for the works mentioned above, this is the case in \cite{erdosyau, erdoseng, DFP}, only in \cite{DF} there is a limiting Markovian density matrix).    A heuristic derivation of a translation invariant Lindblad equation has been advocated recently in  \cite{Vacchini}, see \cite{VH} for a review.

We show that
under the right conditions, the solution of the quantum master equation  behaves diffusively, exactly the
same result as for its classical counterpart: the spatial probability density tends to a Gaussian, after rescaling the position as $x^2 \sim t$ and subtracting a possible
systematic drift. We will also give some counterexamples to that
result thereby
making it less intuitive from a particular point of view.  The questions and the applied techniques
are however quite similar to what has been studied starting from classical Boltzmann-type
equations, see e.g. \cite{bricmontkupiainen,olla,spohnagain}.  A quantum example that
is  very related to ours is in \cite{shenker}. Also, the diffusive limit for another quantum master equation is studied in~\cite{C}.

Intuitively, the dynamics we consider for the particle describes a free ballistic motion which gets interrupted by scattering events with a background fluid in which  momentum  is transfered to the particle. The background fluid is homogeneous and the interaction with the particle is translation invariant.  The intuition for the  particle-environment interaction as occurring through scattering events breaks down somewhat, since, in general, the environment also induces `spatial jumps' for the particle.   This jumping is apparent in a contribution to the diffusion rate which is not driven by the ballistic motion. \\

The next section presents our model and gives a statement of the main result in Theorem \ref{Main}. In Section \ref{disc}, we discuss some examples and a classical analogue. The rest of the paper contains the proof of the main theorem.

\section{Model and results}\label{mod}

\subsection{A translation covariant semigroup}

For the open system dynamics which we consider, the state of the particle at a fixed time $t\geq 0$ is represented by a density matrix $\rho_{t}$ in $ \caB_{1}(\ell^2(\bbZ^d))$, where  $\caB_{1}(\ell^2(\bbZ^d))$ is the space of trace class operators over the Hilbert space $\ell^2(\bbZ^d)$.  To be density matrices the $\rho_{t}$'s must also satisfy $\Tr[\rho_{t}]=1$ and have non-negative eigenvalues.  
The state $\rho_{t}$ evolves  from an initial state $\rho$ as $\Phi_{t}(\rho)=\rho_{t}$ for a family of dynamical maps  $\Phi_{t}:\caB_{1}(\ell^2(\bbZ^d))\to \caB_{1}(\ell^2(\bbZ^d))$, $t\geq 0$ which are  norm-continuous, form a semigroup $\Phi_{t}\Phi_{s}=\Phi_{t+s}$, preserve trace:  $\Tr \Phi_{t}(\rho)= \Tr \rho$ , and are completely positive, see e.g.\ \cite{AF}. 
By~\cite{Lindblad}, a semigroup $\Phi_{t}$ with these properties  must satisfy
\begin{equation}\label{lind}
\frac{d}{dt}\Phi_{t}(\rho)=L(\Phi_{t}(\rho)) \end{equation} where
$ L$ can be written in the \emph{Lindblad} form \beq \label{Eqn:
main repeated} L (\rho) =  -\i [H, \rho] +  \Psi(\rho) -\frac{1}{2} \{ \Psi^{*}(I),
\rho \}, \qquad \rho \in \caB_{1}(\ell^2(\lat)) \eeq in which $H$ is a
self-adjoint operator on $\ell^2(\lat)$, $\Psi$ is a completely positive map acting on $\caB_1(\ell^2(\lat))$, $\Psi^*$  is its adjoint, acting on $\caB(\ell^2(\lat))$, and $I$ is the identity operator on $\ell^2(\bbZ^d)$.   Both $H$ and $\Psi$ are bounded, and their forms have further restrictions discussed below when the $\Phi_{t}$'s have an additional spatial symmetry corresponding to an homogeneous environment.     

To discuss the spatial symmetry that we assume and its consequences, we must define some operators associated with the position and momentum of the particle.   On the Hilbert space  $\ell^2(\bbZ^d)$ we
define the translation operators $\tau_{y}, y \in \lat$,  and the
(vector-valued) position operator $X$ as
\[
(\tau_{y}f)(x) = f(x+y), \qquad    (X_j f)(x)=   x_j f(x) \,\; \mbox{ for
} \, f\in  \ell^2(\bbZ^d), \, x \in \lat .
 \]
 Here, and in what follows, the subscript $j=1,\ldots,d$ refers to the components in $\lat$ or $\tor$.
 We will often consider the space $\Hi$ in its dual representation, i.e.,
  as $L^2(\bbT^d) $ where $\bbT^d$ is
  identified with $[-\pi,\pi]^d$.
   For $g \in L^2(\bbT^d)$, we define
the vector of `momentum' operators $\Om= (\Om_j)$ as multiplication
   by $k \in \bbT^d$, i.e.,
\beq \Om_{j} g\,(k)= k_{j}\,g(k), \quad  k \in [-\pi,\pi]^d \eeq
Although the $\Om_{j}$'s are well-defined as bounded operators,
they do not satisfy $[X_{j},\Om_{j}]=\i$, nor does $\Om$ generate
the translations $\tau_x$.  Trace class operators $\rho$ are Hilbert-Schmidt and thus have well-defined square integrable kernels $\rho(x_1,x_2):\bbZ^{2d}\rightarrow \bbC $ and $\rho(k_1,k_2): \bbT^{2d}\rightarrow \bbC $, which are related by the Fourier transform 
\beq
\rho_t(k_1,k_2) := \frac{1}{(2\pi)^d} \sum_{x_1, x_2}   e^{-\i (x_1 k_1-x_2 k_2)}   \rho_t(x_1,x_2), \qquad  k_1,k_2 \in \tor.
\eeq 
We refer to $\rho(x_1,x_2)$ and $\rho(k_1,k_2)$ as the position and momentum representation of the state $\rho$.

We demand that the semigroup $\Phi_{t} $  be translation covariant
\[\Phi_{t}(\tau_{x}^{*}\rho
\tau_{x})=\tau^{*}_{x}\Phi_{t}(\rho)\tau_{x},\qquad  x \in \lat . \]
By \cite{Holevo}, this implies  that one can choose $H = H(\Omega)$,  a bounded
function of the vector of momentum operators $\Omega$ and that
  $\Psi$ has what we refer to as the Holevo-form:
\begin{align}\label{Holevo}
\Psi(\rho)= \int_{\bbT^{d}}\d \nu(\theta)\, e^{ \i  \theta X}\mathbf{M}_{\theta}
( \rho )e^{-\i  \theta X},
\end{align}
where $\nu$ is a positive finite measure on the
$d-$dimensional torus $\bbT^{d}$, and the maps $\mathbf{M}_{\theta}$ are completely positive and act as multiplication in the momentum representation
\begin{align}\label{Mult}
\mathbf{M}_{\theta}(\rho)(k_{1},\,k_{2})=M_{\theta}(k_{1},\,k_{2})\rho(k_{1},\,k_{2})
\end{align}
 for some functions $M_{\theta}(k_{1},\,k_{2}):\bbT^{2d}\rightarrow \bbC$.  Note that the vector of position operators $X$ generates torus translations in the variable $k$, i.e., $e^{ -\i  \theta X}\psi(k)= \psi(k+\theta)$, for $\psi\in L^{2}(\bbT^{d})$ and thus, in momentum representation,
 \beq \label{Shift} \big(e^{ \i  \theta X}\rho e^{-\i  \theta X})(k_{1},k_{2})= \rho(k_{1}-\theta,k_{2}-\theta).\eeq
Boundedness of the functions $H(k)$ and $\int_{\tor}d\nu_{\theta}M_{\theta}(k,\,k)$ in $k\in \tor$ is equivalent to the translation covariant semigroup being norm-continuous.  Since the Hamiltonian $H(\Omega)$ is a function of the vector of momentum operators $\Omega$, the kinetic motion of the particle is driven by its momentum.
  Intuitively,  the map $\mathbf{M}_{\theta}$ encodes the frequency and nature of the scatterings which the particle receives through collisions with the reservoir that result in a momentum transfer $\theta$.   As mentioned in  Section \ref{sec: assumptions}, the maps $\mathbf{M}_{\theta}$  also generate spatial jumps for the particle which contribute to the diffusion constant (see the end of Section~\ref{sec: symmetries}). 

There is a crucial decomposition of the dynamics in  the momentum representation as a consequence of the translation symmetry. It is useful to
change
  variables in the momentum representation and to write
\beq \label{Fiber: first} [\rho]_p(k):=  \rho(k- \frac{p}{2},k+\frac{p}{2}) , \qquad   k,p \in \tor \eeq ,
 where we will think of  $[\rho]_p$ as  fibers of the density matrix $\rho$, indexed by $p \in \tor$.
   The equation \eqref{Fiber:
first}  defines a map
$$[\cdot]_{p}:\caB_{1}(\ell^2(\bbZ^{d}))\rightarrow L^{1}(\tor).$$
We will study these maps with more care in Lemma \ref{lem: fiber}. 
In particular, our conditions on the initial state $\rho$ will ensure that the function $p \mapsto [\rho]_{p}$ can  be chosen in $\caC^2(\tor, L^1(\tor))$.
Due to the translation symmetry, the dynamics gives rise to an autonomous evolution for each fiber $[\rho_{t}]_{p}$.  This can be seen from (\ref{Holevo})-(\ref{Shift}) and the fact that both $\Psi^{*}(I)$ and $H$ are functions of the momentum operator and thus act as multiplication in momentum space (see also below under \eqref{46}).  In particular, the momentum distribution for the particle, given by the diagonal $[\rho]_{0}(k)=\rho(k,k)$ in the momentum representation, undergoes a classical Markovian evolution.  \\

 The position
 distribution for the particle is found on the diagonal in the position representation, $\rho_{t}(x,x)$, which is itself
  not Markovian.
The main and mathematical result of the paper
 is the identification of natural assumptions on the Hamiltonian $H$ and the maps $\mathbf{M}_{\theta}$ under which the measure $\mu_{t}$, defined by
 \[
 \mu_{t}(R)=
 \sum_{x\in \sqrt{t}R-tv}\rho_{t}(x,x)
 \]
  for a drift velocity $v\in \bbR^{d}$ and for an  arbitrary Borel set $R\subset \bbR^{d}$,
  converges in distribution to a Gaussian law.   Our mathematical assumptions on the kernels
 $M_\theta(k_1,k_2)$ basically express a certain locality in the spatial jumps and  a sufficiently smooth relaxation for the momentum distribution.

\subsection{Assumptions}\label{sec: assumptions}

There are basically two sets of assumptions, one having to do with
the spatial locality of the dynamics, and the other with the
dissipativity.
We  formulate these
conditions below, and we discuss them more closely in Section \ref{disc} after having
stated the main
result of the paper.

The locality can  be formulated in terms of the completely
positive maps $\mathbf{M}_{\theta} $, the measure $\d \nu(\cdot)$  and the dispersion law $H$.  Note that there is
a slight arbitrariness in choosing $\d \nu$ and $\mathbf{M}_\theta$
since only the measures $\d \nu(\theta){M}_\theta(k_{1},k_{2})$
enter in the definition of $\Psi$.

\begin{assumption}\label{as1}[Locality]
We assume that  the completely
positive maps $\mathbf{M}_{\theta} $ are defined by the kernels ${M}_{\theta}(k_{1},k_{2})$ as
 \beq \label{Who}
 (\mathbf{M}_{\theta}\rho)(k_1,k_2)=
{M}_{\theta}(k_{1},k_{2})\,\rho(k_{1},k_{2}) \eeq
where the function $M_{\theta}(k_1,k_2)$ is twice continuously
differentiable in both $k_1,k_2$.  Moreover, the matrix of derivatives $D^{2}M_{\theta}(k_{1},k_{2})$ is uniformly bounded in $\theta,k_1,k_2\in \tor$, and the family of functions $D^{2}M_{\theta}, \theta\in \tor$ is equicontinuous.   
The function $H$ is assumed to be twice continuously differentiable.  Finally, we assume that  $ \d \nu(\cdot) $ in
\eqref{Holevo}  is a probability measure.
\end{assumption}

It is instructive to examine the map $\Psi$ in position representation.  Let $\tilde{M}_{\theta}:\bbZ^{2d}\rightarrow \bbC$ be the double Fourier transform 
$$\tilde{M}_{\theta}(x_{1},x_{2})=\frac{1}{(2\pi)^{d} }\int_{\bbT^{2d}}dk_{1}dk_{2}\,e^{-\i k_{1}x_{1}+\i k_{2}x_{2}} M_{\theta}(k_{1},k_{2}).$$
The operation of $\Psi$ in the position representation has the form
\begin{eqnarray}
\Psi(\rho)(x_{1},x_{2})&=& \int_{\tor}d\nu(\theta)e^{\i(x_{1}-x_{2})\theta}\sum_{y_{1},y_{2}\in \bbZ^{d} } \tilde{M}_{\theta}(x_{1}-y_{1},x_{2}-y_{2})\rho(y_{1},y_{2}) \nonumber \\ \label{SpatJumps} &=&  \sum_{y_{1}, y_{2}\in \bbZ^{d}}
N(x_{1},y_{1},y_{2},x_{2})\rho(y_{1},y_{2}) ,  
\end{eqnarray}
where the second equality determines the values of the kernel $N:\bbZ^{4d}\rightarrow \bbC$.  It is apparent from the form above that the noise term $\Psi$  generates some spatial jumping unless $N(x_{1},y_{1},y_{2},x_{2})$ vanishes whenever $x_{1} \neq y_{1}$ or $y_{2} \neq x_{2}$.  The translation covariance of $\Psi$ is expressed through the equality
\[
N(x_{1},y_{1},y_{2},x_{2})= N(x_{1}+z,y_{1}+z,y_{2}+z,x_{2}+z),
\qquad \textrm{for all}  \, z \in \bbZ^d. \]
The conditions on the $\mathbf{M}_{\theta}$ in Assumption~\ref{as1} can be replaced by the assumption   
\[
 \sup_{y_{1},y_{2}\in \bbZ^{d}  }  \sum_{x_{1}, x_{2}\in \bbZ^{d}}
\big((x_{1}-y_{1})^{2}+(x_{2}-y_{2})^{2}\big) |
N(x_{1},y_{1},y_{2},x_{2})|<\infty.
\]
to obtain our results.  This locality condition can be compared with asking for finite variance in
the jumps in a random walk, and it
is supposed to exclude superdiffusive behavior.\\

We now come to the dissipativity. To formulate that
conveniently, we introduce the Markov generator $A$ on $L^1(\tor)$
\beq \label{ir1}
 (A f)(k) = \int_{\tor} \d \nu(\theta)\,r(k,k-\theta )\,
f(k-\theta)  -   \int_{\tor} \d \nu(\theta)\, r(k+\theta,k) f(k)
\eeq with the  transition rates $r(k,k')$  defined by,
\beq \label{def: rates from m}
  r(k+\theta,k):= M_\theta(k,k) \geq 0.
  \eeq
  The measure $\d \nu(\theta)  r(k+\theta,k)$ gives the probability density per unit time of jumping to the state $k+\theta$, conditioned on being in the state $k$.  As before, the parameter $\theta$ plays the role of the momentum  transfer.

Remember that $r(\cdot,\cdot)$ is a $\caC^1$-function by Assumption \ref{as1}.
Obviously, one has $\norm A f \norm_1 \leq c  \norm f \norm_1$ for  any
 $f \in\caC(\tor) $ and $c:= 2 \norm r(\cdot,\cdot)\norm_{\infty}$, and hence $A$ is the bounded generator of a contractive (and positive)
  semigroup on $L^1(\tor)$, see the connection with the Hille-Yosida theory in
e.g.~\cite{Liggett}). 
We now discuss dissipativity, which refers to ergodic
properties. 

\begin{assumption}\label{ass: irreducibility}[Dissipativity]
We assume
\begin{enumerate}
\item {$A$ has a simple eigenvalue $0$ with eigenvector $\caP \in L^1(\tor)$, normalized such that $\int_{\tor} \d k \caP(k)=1 $}
\item  The eigenvalue $0$ is separated from the  rest of the $L^1$-spectrum by a gap $b_A$,  \[   b_A:=    -\sup \mathrm{Re}  \left(\mathrm{spec}(A) \setminus \{ 0 \} \right)  > 0 \]
\end{enumerate}
\end{assumption}

The above assumptions  guarantee that the semigroup generated by $A$, i.e., $\e^{tA}$, relaxes exponentially fast to the stationary distribution $\caP$. For future use, we let $Y_t$ stand for the Markov process on $\tor$ generated by $A$ and started from $\caP$.
Using
standard techniques for Markov processes on compact spaces,
constructive conditions are available for guaranteeing the
Assumption \ref{ass: irreducibility} in terms of  $r(k,k')$ and
$\nu$.  While the above assumptions are natural ergodicity or gap-assumptions, in fact
for our result we need less. In particular, the exponential
relaxation is not strictly necessary. We will however not describe
that.

The following standard construction will be useful in the statement of one of our results.
Consider the scalar product
\[
\langle f,g \rangle_\caP := \int  \d k  \caP(k)   \overline{f(k)} g(k), \qquad  f,g \in \caC(\tor)
\]
and let $\caH_\caP$  stand for the Hilbert space which is the
completion of $\caC(\tor)$ with respect to the scalar product
$\langle \cdot, \cdot \rangle_\caP$. Define  the quadratic form
$A_\caP$ by \beq \label{def: acap} \langle    f,  A_\caP g
\rangle_\caP  = \int  \d k  \caP(k)   \overline{(A^*f)(k)}  g(k),
\qquad  f,g \in \caC(\tor) \eeq
where $A^*$ is the adjoint of $A$ acting on $L^{\infty}(\tor)$.\\

The operator $A$ appears naturally in a perturbation set-up around
the zero fiber.  In fact, the evolution on the zero fiber is the
Markov process generated by $A$.

We finally ask some properties that appear directly linked with
the notion of diffusion.  We have of course already that the
particle must be sufficiently localized since it is described by a density matrix
$\rho \in \caB_{1}(\ell^2(\bbZ^d))$,
  but we also ask that the first two moments are well-defined in the following way
\beq \label{cond: regularity initial} X_j\rho, \quad    X_i \rho X_j , \quad  X_i X_j\rho \quad\textrm{are in}\,\, \caB_{1}(\ell^2(\bbZ^d)), \qquad  \textrm{for}\, i,j=1, \ldots, d. \eeq
Products of bounded and unbounded operators such as in~(\ref{cond: regularity initial}) are to be understood as kernels of sesquilinear forms with densely defined domains.  For example, $X_{j}\rho\in \caB(\ell^2(\bbZ^d)) $ (which is implied since $\caB_1 \subset \caB$)  means that  $\mathit{b}(f,g):=\langle X_{j}f, \rho g\rangle$ satisfies $\str \mathit{b}(f,g) \str  < C \norm f \norm_2 \norm g \norm_2$ for  $(f,g)\in \Dom(X_{j})\times \ell^{2}(\bbZ^{d})$ and some $C<\infty$, and thus the quadratic form $b(\cdot,\cdot)$ is extendable to $\ell^{2}(\bbZ^{d}) \otimes \ell^{2}(\bbZ^{d})$.   In particular, note that by the boundedness of the form $b(f,g)$ and by the definition of the domain of the self-adjoint operator $X_{j}$, we have that
\beq \label{Give}
\rho  g \in \Dom(X_{j}), \qquad  \textrm{for any} \, \,   g \in \ell^{2}(\bbZ^{d}),
\eeq
so that the  product $X_{j}\rho$ makes sense.

We write in general  $\rho_{t}$ for the solution of the equation
\eqref{lind} with initial condition $\rho_0 \in \caB_1$. We will
show that there is a $v \in \bbR^d$ such that \beq v= \lim_{t
\nearrow \infty} \frac{1}{t} \sum_{x \in \lat}  x \rho_t(x,x) \eeq That can be
strengthened to a weak law of large numbers.  One can obviously
force $v=0$ by requiring some additional symmetries. Getting our
results does however not depend on these extra requirements.  In particular,
equilibrium conditions such as detailed balance are mostly
irrelevant for the diffusive behavior around the drift, except
when one wants for example to relate the diffusion constant to the
mobility in linear response theory.

\subsection{Result}

We define $T^{(1)}$ and $T^{(2)}$ as, respectively, a  vector and $d\times d$ matrix of operators on $L^1(\tor)$, by
\beq \label{FirstOrder} (T^{(1)}f)(k)=-\i(\nabla H)(k)f(k)+\int_{\bbT^{d}}
d\nu(\theta)\, m_{\theta}^{(1)}(k-\theta)f(k-\theta),  \eeq
where
$m_{\theta}^{(1)}(k)=
-i\textup{Im}\big(\nabla_{1}M_{\theta}(k,k)\big)$
(and $\nabla_{1}$ and $\nabla_{2}$ are the gradients with respect
to the first and second variables of $M_{\theta}(k_{1},k_{2})$),
and
 \beq\label{t2}
(T^{(2)}f)(k)=\int_{\bbT^{d}}d\nu(\theta) \bigg(
(m_{\theta}^{(3)}-m_{\theta}^{(2)}) (k-\theta)f(k-\theta)
-m_{\theta}^{(3)}(k)  f(k)  \bigg)\eeq where
\[
\big(m_{\theta}^{(2)}(k)\big)_{(i,j)} =
\frac{1}{4}\big[\big(\nabla_{1}\nabla_{2}M_{\theta}(k,k)\big)_{(i,j)}+
\big(\nabla_{1}\nabla_{2}M_{\theta}(k,k)\big)_{(j,i)}\big]\text{,\quad and}
\]
\[ \big(m_{\theta}^{(3)}(k)\big)_{(i,j)}=\frac{1}{2}\big[\big(\nabla^{2}_{1}+\nabla^{2}_{2})M_{\theta}(k,k)\big)_{(i,j)}+\big(\nabla^{2}_{1}+\nabla^{2}_{2})M_{\theta}(k,k)\big)_{(j,i)} \big]  .\]

We let $P_{0}$ be the spectral projection corresponding to the $0$
eigenvalue of $A$, the Markov generator defined in \eqref{ir1},
and take $S$ the reduced resolvent of $A$ at the eigenvalue $0$,
i.e.,\ the solution of
\[
S (0-A) =    (0-A) S = 1-P_0 , \qquad  S P_0=P_0 S =0.
\]
Finally, recall that  $\mathcal{P}$ is the eigenvector corresponding to the eigenvalue $0$. Hence $\caP(k) \d k$ is  the stationary probability
measure on the torus.  In general the projection $P_{0}$ is non-orthogonal and has the form $P_{0}=| \mathcal{P}\rangle \langle 1_{\bbT^{d} }|$, where $1_{\bbT^{d}}$ is the constant function on $\bbT^{d}$ with value $1$.  We use the notation $\langle g, f \rangle:= \int_{\tor} \d k \,  \overline{g(k)} f(k)$ for the pairing between $f \in L^1(\tor)$ and $g \in L^\infty(\tor)$.
To the time-evolved density matrix $\rho_t$, we associate  the measures $\mu_{t}$ on $\bbR^d$, defined by
\beq   \label{def: measure mu}
\mu_{t}(R):= \Tr[1_{\sqrt{t}R-vt }(X)\,\rho_{t}],\qquad  \textrm{for a Borel set}  \, \, R \subset \bbR^d\eeq
where $1_{\sqrt{t} R-vt }(X)$ is the spectral projection of the
vector of position operators $X$ on the set
$\sqrt{t}R-vt\subset \bbR^{d}$.

\begin{theorem}\label{Main}
Take Assumptions \ref{as1} and \ref{ass: irreducibility}  and let  the initial density matrix $\rho_0 \in \caB_1(\ell^2(\lat))$  satisfy \eqref{cond:
regularity initial}.
The limiting velocity exists and equals
\beq \label{def: velocity}
v := \lim_{t
\nearrow \infty} \frac{1}{t}   \sum_{x \in \lat}  x \rho_t(x,x) = -\i \langle 1_{\tor},  T^{(1)} \caP  \rangle
\eeq
The measure
 $\mu_{t}$, defined as in \eqref{def: measure mu}  with $v$ as in \eqref{def: velocity},
converges, as  $t\rightarrow \infty$, in distribution to a Gaussian with covariance matrix $\sigma=\beta+\beta^{\dag}$, where $\beta^{\dag}$ is the transpose of the matrix $\beta$, given by
\begin{equation}\label{ginger}
\beta :=-\langle 1_{\tor}, \big(T^{(2)}-T^{(1)}ST^{(1)}\big) \mathcal{P}\rangle.
\end{equation}
where the operators  $T^2$ and $T^1$ as defined above.

The truncated second moments converge to the covariance matrix $\sigma$, i.e.\
\begin{equation}\label{ginger2}
  \lim_{t \nearrow \infty}   \sum_{x \in \lat}  \rho_t(x,x)   (x_i-tv_i)(x_j-tv_j) =     \sigma(i,j)
\end{equation}
\end{theorem}
 Equation  \eqref{ginger2}  tells us that  the covariance matrix $\sigma$ can be interpreted as the  diffusion tensor.   In the case that  $T^{(1)}=-\i \nabla H(\Omega) $ and $T^{(2)}=0$, it reduces to another matrix that below we call $\alpha$.  In the proposition below, the matrices $\sigma$ and $\alpha$ are ``non-negative" in the sense of real valued vectors:  $\forall (v\in \bbR^{d})$, $(v,\sigma\,v),(v,\alpha\,v)\geq 0$.
 \begin{proposition}\label{pos}
Consider the covariance matrix $\sigma$ as above.
\begin{enumerate}
\item $\sigma$ is non-negative.
\item Define the (vector) function
\beq \zeta:= \nabla H - \langle   \nabla H, \caP \rangle
\eeq
on $\tor$.
The real-valued $ d \times d-$matrix  $\alpha$ with entries
\beq \label{Pineapple}
\alpha(i,j)=
\frac{1}{2}\int_{0}^{\infty}dt\,
\bbE_{\mathcal{P}} [(\zeta(Y_{t}))_{i} (\zeta(Y_{0}))_{j}+(\zeta(Y_{t}))_{j} (\zeta(Y_{0}))_{i}   ],
\eeq
is non-negative.
(As in Section \ref{sec: assumptions}, $Y_{t}$ is the stationary Markov process on $\bbT^{d}$ generated by $A$ and started from $\caP(k)\d k$, cf.~(\ref{ir1})).

\item
Assume that for all $w \in \bbR^d$, the function $ k \mapsto (w, \zeta(k))$ does not vanish identically (we write $(\cdot,\cdot)$ for the scalar product on $\bbR^d$). That is, all components of the velocity $\nabla H$ can fluctuate.  Assume in addition that  $A_\caP$, defined as a quadratic form in \eqref{def: acap}, extends to a bounded and sectorial operator on $\caH_\caP$. Sectoriality \cite{Kato} means that there is  $0<\ga <\infty $ such that
\beq
 \str  \langle f , \mathrm{Im}(A_\caP)  f \rangle_\caP \str   \leq  - \ga    \langle f , \mathrm{Re} (A_\caP)  f \rangle_\caP , \qquad  \textrm{for all} \, f \in \caH_\caP.
\eeq
Then the matrix $\al$ is strictly positive (i.e., it has strictly positive eigenvalues).
\end{enumerate}
\end{proposition}

\section{Discussion}\label{disc}

\subsection{Properties of the drift and diffusion constant} \label{sec: symmetries}
We can erase  drift terms and simplify the diffusion matrix $\si$ by imposing further
symmetries.\\
Define the linear and antilinear maps $U$ and $V$ acting on
$f \in \ell^{2}(\bbZ^{d})$ as
\[
(Uf)(x)=f(-x)  \quad \text{and} \quad  (Vf)(x)=\overline{f(x)}
\]
If we assume the space-inversion symmetry \beq \label{drift}
U\Phi_{t}(\rho)U^{-1}=\Phi_{t}(U\rho U^{-1}),\hspace{.5cm} \rho\in
\caB_{1}(\ell^{2}(\bbZ^{d})) \eeq (or on the generator $L$ of the
semigroup $\Phi_{t}$), we have no drift i.e. \beq 0=v= \lim_{t
\nearrow \infty} \frac{1}{t} \sum_{x} x \rho_t(x,x)  \eeq


%

If we assume the symmetry (for the noise term)
\beq \label{Symmetry}
 W\Psi(\rho)W^{-1}=\Psi(W\rho W^{-1}),   \qquad  W=UV,
\eeq
 then $M_{\theta}(\cdot,\cdot)$ is real, consequently the function $m_{\theta}^{(1)}$ in~(\ref{FirstOrder}) is zero and $T^{(1)}=-i\nabla H $.  This means that the drift is given by $v = \langle \nabla H, \caP \rangle$ and the diffusion matrix $\si$ simplifies to the form 
\begin{multline}  \label{Matrix}
\sigma(i,j)=\int_{\tor}dk\,  \int_{\tor} d\nu(\theta) \,
\mathcal{P}(k)\,( [\nabla_{1}\nabla_{2}M_{\theta}]_{(i,j)}(k,k)+[\nabla_{1}\nabla_{2}M_{\theta}]_{(j,i)}(k,k)) \\+\int_{0}^{\infty}dt\,
\bbE_{\mathcal{P}} [(\zeta(Y_{t}))_{i} (\zeta(Y_{0}))_{j}+(\zeta(Y_{t}))_{j} (\zeta(Y_{0}))_{i}   ],
\end{multline}
where the rightmost expression is defined as in~(\ref{Pineapple}). Notice that the vanishing of $m^{(3)}_{\theta}$ from these expressions is a general feature that does not depend on  \eqref{Symmetry}.  The second term in \eqref{Matrix} is the diffusion constant of a classical Boltzmann equation with momentum evolution given  by the Markov process $Y_t$.
The form makes the contribution of spatial jumps to the diffusion matrix $\sigma$ clear.   The first term on the RHS of \eqref{Matrix}  depends on the noise but not on the dispersion relation $\nabla H$.  This part of the diffusion must arise from the noise-induced spatial jumps related to the position representation~(\ref{SpatJumps}).  A special case where this term disappears is discussed in Section~\ref{Sec: Examples}.   

We sketch why this first term is a non-negative matrix. 
  Since $\mathbf{M}_\theta$ is a completely positive map on $\caB_1(\ell^2(\lat))$, for each $\theta$, it follows that $(v, [\nabla_{1}\nabla_{2}M_{\theta}]v)$ is the integral kernel of a positive operator on $L^{2}(\bbT^{d})$.   This follows since $ \mathbf{M}_{\theta}(|f\rangle \langle f|)$ is a positive bounded operator for any $f\in L^{2}(\bbT^{d})$
and thus for any $g\in L^{2}(\bbT^{d})$
\begin{align}\label{Tisdale}
0\leq  \langle g,\, \mathbf{M}_{\theta}(|f\rangle \langle f|) g \rangle= \int_{\mathbb{T}^{2d}} M_{\theta}(k_{1},k_{2})\bar{h}(k_{1}) h(k_{2})  \quad \quad h(k)=\bar{f}(k)g(k).       
\end{align}
Through choice of $f,g$, the functions $h(k)$ can be formed into arbitrary elements in $L^{1}(\bbT^{d})$. Choose $h=\big(v\cdot \nabla h_{0}\big)(k)$ for any once continuously differentiable function $h_{0}(k)$.  Using integration by parts twice, we have  
$$0\leq  \int_{\mathbb{T}^{2d} } dk\, \bar{h}_{0}(k_{1})h_{0}(k_{2})(v, [\nabla_{1}\nabla_{2}M_{\theta}]v) (k_{1},k_{2}).  $$
The positivity can be extended to all $h_{0}\in L^{2}(\bbT^{d})$, since $M_{\theta}(k_{1},k_{2})$ is twice continuously differentiable and thus $(v, [\nabla_{1}\nabla_{2}M_{\theta}]v) (k_{1},k_{2})$ determines a  Hilbert-Schmidt operator.  The diagonal entries in the integral kernel for a positive operator are non-negative, so     
$$(v, \int_{\tor} dk\, d\nu(\theta) \,
\mathcal{P}(k)\, [\nabla_{1}\nabla_{2}M_{\theta}] (k,k)v)$$
is thus non-negative as claimed.    Symmetrizing $[\nabla_{1}\nabla_{2}M_{\theta}]_{(i,j)}(k,k)$ in the above expression makes no difference in the evaluation of expressions $(v,\sigma v)$ for $v\in \bbR^{d}$ but ensures that $\sigma $ is a symmetric and real-valued diffusion matrix.

\subsection{Idea of proof: perturbation theory}\label{Sec: Idea}
The main feature of our translation invariant models is that the
evolution generated by \eqref{lind} can be decomposed along
the fibers~(\ref{Fiber: first}), i.e., one can write $[ L \rho ]_p  = L_p [\rho]_p$  for some operators $L_p$ and the fibers of the density matrix, $[\rho_{t}]_{p}$,  obey
the differential equation \beq\label{46}
\frac{d}{dt}[\rho_{t}]_{p}=   L_p [\rho]_p . \eeq
The expression for  $L_{p}$   can be determined as a quadratic
form through a trace formula: for $F\in L^{\infty}(\bbT^{d})$,
 \beq \label{Explicit}
 \Tr[ L^{*}\big( \e^{\i \frac{p}{2} X}
 F(\Omega)\e^{\i \frac{p}{2} X}  \big) \rho ] =
 \langle F, L_{p}[\rho]_{p}\rangle=\int_{\bbT^{d}}dk\, F(k)\, L_{p}[\rho]_{p}(k).
 \eeq
    A simple computation shows that
       $L_{p}= -\i h_p + \varphi_p$ with
\begin{eqnarray}\label{Fiber: again}
 ( h_{p} f)(k) &:=& (H(k-\frac{p}{2})-H(k+\frac{p}{2})) f(k) \\
 (\varphi_p f)(k)&:= \label{Fiber: again2} &\int_{\bbT^{d}} d\nu(\theta) \big( r_{p}(k-\theta,k) f(k-\theta)-\bar{r}_{p}(k,k+\theta) f(k)\big),
\end{eqnarray}
where
$r_{p}(k,k^{\prime})=M_{k^{\prime}-k}(k-\frac{p}{2},k+\frac{p}{2})$. Note that $r_{p=0}(k,k^{\prime})$ was simply called $r(k,k^{\prime})$ in \eqref{def: rates from m}.  By the inequality~(\ref{Tisdale}) for a sequence of $h_{n}=h^{\prime}_{n}$ approaching a $\delta$-function, it follows that the values  $r(k,k^{\prime})$ are nonnegative.

 Since only the
fibers around $p=0$ determine the diffusive behavior, this
suggests using a perturbation argument to capture the essential
dynamical properties of those fibers.  Under our assumptions on $A= L_{0}$, it has a gap between the zero
eigenvalue corresponding to the stationary distribution
$\mathcal{P}$ and the rest of the spectrum
which has strictly negative real part.    Hence sufficiently small bounded perturbations of $L_{0}$
keep the gap open.\\

  The following proposition gives a condition on $\rho$
   such that the function $p \mapsto [\rho]_{p}$ is twice differentiable. This assures the existence of the first two moments.

 \begin{proposition}\label{lem: fiber}
Assume that a density matrix  $\rho\in \caB_{1}(\ell^2(\lat))$ satisfies
  \beq \label{eq: ass moments exist} X_j \rho,\quad  X_i \rho X_j,\quad  X_i X_j \rho \quad \textrm{are in} \,\,\caB_{1}(\ell^2(\lat)) \,\,\textrm{for}\,\, i,j=1, \ldots, d. \eeq
  (as in \eqref{cond: regularity initial}), then the function
    $\bbT^{d}\rightarrow L^{1}(\bbT^{d}):  p \mapsto [\rho]_{p}$ can be chosen to be twice continuously differentiable.
\end{proposition}

\subsection{Examples}\label{Sec: Examples}
We mention a certain subclass of models satisfying
 the
 conditions for~(\ref{Main}) and  another class of models that do not satisfy our conditions and that possibly
show a non-diffusive (e.g.  super-diffusive) behavior.

We  consider first the case when the completely positive maps
$\mathbf{M}_{\theta}$ in \eqref{Holevo} act identically $\mathbf{M}_{\theta} \rho=\rho$, such that $\Psi$ becomes
\begin{align}\label{Faluja}
\Psi(\rho)=\int_{\bbT^{d}} d\nu(\theta)\, e^{\i  \theta X} \rho e^{ -\i  \theta X}
\end{align}
The map $\Psi$ operates multiplicatively in the position representation
$$\Psi(\rho)(x,y)= \varphi(x-y)\rho(x,y)$$
where $\varphi$ is the Fourier transform of the measure $\nu$. 
A noise of the type~(\ref{Faluja}) has appeared as a tight-binding approximation for the modeling of a low energy atom in a periodic potential~\cite{Kolovsky}.  Also, it is similar to the noise term for Gallis-Flemming
dynamics~\cite{Gallis} for a quantum particle in $\bbR^{3}$ that has
been frequently discussed in the decoherence
literature~\cite{Alicki,Sipe}.  The derivation in~\cite{Gallis} starts from a scattering framework and considers a limiting regime where the mass of the particle is much larger than the mass of particles in a background gas.  The jump rates $d\nu(k-k^\prime)r(k,k^\prime)=d\nu(k-k^{\prime}) $ from the zero-fiber process~(\ref{ir1}) only depend on the difference between the starting momentum $k^\prime$ and landing momentum $k$.  The dynamics can thus be described as frictionless, and the stationary density $\mathcal{P}$ of the zero-fiber  Markov process is the uniform distribution.     Since
$\nabla_{1}\nabla_{2}M_{\theta}(k_{1},k_{2})=0$, the diffusion
constant $\sigma$ takes the
familiar form of the second term in~(\ref{Matrix}). That is, the diffusion matrix reduces to $\al$, as defined in Proposition \ref{pos}.\\

Thinking about adding spatial jumps we arrive at models where the
noise $\Psi$ resembles a simple symmetric random walk.  Yet, that
easily breaks Assumption~\ref{ass: irreducibility} Statement 2, and the model can become super-diffusive when
the kinetic term is included.  As an illustration, we consider a
one-dimensional model. Let $\Psi$ have the form
$$\Psi(\rho)(x_{1},x_{2}) = \sum_{y_{1},y_{2}\in \bbZ}N(x_1,y_1,y_2,x_2)\,\rho(y_1,y_2),$$
where $N(x_{1},y_{1},y_{2},x_{2})$  is determined by a function
$r(x), x\in \bbZ^d,$ with a positive Fourier transform through the
equation
\[
N(x_1,y_1,y_2,x_2) = r(x_1-x_2)\,
\chi[|x_1-y_1|=1]\,\chi[|x_2-y_2|=1],
\]
where $\chi[\cdot]$ is the indicator  (1 or 0).  In that case, the measure
$\d\nu(\cdot)$ can be taken to be Lebesgue measure and the
functions $M_{\theta}(k_{1},k_{2})$ are determined by

\begin{eqnarray}
M_{\theta}(k_{1},k_{2}) &=&4\,  \hat{r}(\theta)  \,\cos (k_1) \,\cos
(k_{2})
 \nonumber \\
\Psi(\rho)(k_1,k_2) &=& 4\int_{\bbT}d\theta\,\hat{r}(\theta)
\,\cos (k_{1}-\theta) \,  \cos(k_{2}-\theta )\,
\rho(k_{1}-\theta,k_{2}-\theta) \end{eqnarray} The smoothness of
$\hat{r}(\theta)=\frac{1}{(2\pi)^{d}}\sum_{n}e^{\i  n\theta}r(n)$
depends on the decay of $c$.  Notice  that the Markov process of
the zero fiber is generated by
\beq\label{Zabba}
(Af)(k)=
4\int_{\bbT}d\theta\,\hat{r}(\theta)\, \big(\cos
(k-\theta)^{2}f(k-\theta)-\cos(k)^{2}f(k)\big).  
\eeq
That is
describing a Markov process on the torus $\bbT$ where an escape from
the point $k$ occurs with rate $4 \cos^{2}(k) \,(\int_{\bbT}
d\theta \,
 \hat{r}(\theta))$ and  the jump size $\theta$ is
 independent of $k$ and has probability density
    $$\frac{\hat{r}(\theta)}{\int_{\bbT}d\theta\, \hat{r}(\theta)}$$
Assumption~\ref{ass: irreducibility} is now violated in that, for
$k=\pm \frac{\pi}{2}$,
 $$\int_{\bbT} d\theta \, M_{\theta}(k,k)= 4(\int_{\bbT} d\theta\, \hat{r}(\theta)) \cos^{2}(k)=     0$$
This implies that there are degenerate stationary distributions of the form
\beq \label{Delta}
\mathcal{P}_{\lambda}(k)=\lambda \delta \big(\frac{\pi}{2}-k\big)+(1-\lambda)\delta\big(-\frac{\pi}{2}-k\big)
\eeq
for $0\leq\lambda \leq 1$.  The process is thus slow to leave the regions around $k=\pm \frac{\pi}{2}$.  We conjecture that for an arbitrary smooth probability distribution $\caV$ on $\bbT$, $e^{tA}\caV$ will converge in distribution to $\caP_{1/2}$, i.e., to \eqref{Delta} for $\lambda=\frac{1}{2}$.

When the kinetic term $H$ is zero,  we observe that still the
diffusion constant as such keeps making mathematical sense as
$$\sigma = 2\int_{\bbT}dk\,\mathcal{P}_{\frac{1}{2}}(k)(\partial_{1}\partial_{2}M)(k,k)
= 8 \int_{\bbT}d\theta \, \hat{r}(\theta)$$
When  the kinetic term $H$ is non-zero and $H^{\prime}(\pm
\frac{\pi}{2})\neq 0$, we expect that the model exhibits a
behavior closer to being ballistic with a width in position that
grows on the order of $t$ rather than $t^{\frac{1}{2}}$.    The
basic idea is that when the particle attains a momentum close to
$k=\pm \frac{\pi}{2}$, then it tends to move ballistically without interruption with that momentum for long intervals of time.  A classical analog of this process was studied in~\cite[Sec. 4]{Jara} for the case such that $ H^\prime(k\pm \frac{\pi}{2})\approx \pm c|k\pm \frac{\pi}{2}|^{\gamma}$ when $|k\pm \frac{\pi}{2}|\ll 1$ for some $c>0$ and $\gamma \geq 1$. They consider a linear functional  $Y_{t}=\int_{0}^{t}dr\,V(K_{r})$ for a dispersion relation $V:\mathbb{T} \rightarrow \bbR$ (e.g. $V=H^{\prime}$) and Markov process $K_{r}$ whose densities obey the master equation~(\ref{Zabba}) (and for a more general class of jump rates).  They show that $N^{-\frac{\beta}{2} }Y_{Nt} $ converges in law to a limiting process as $N\rightarrow \infty$ for $\beta=\frac{1}{2}$ when $\gamma>1$ and $\beta=1$ when $\gamma=1$.  These correspond to situations where the velocities are small for the regions of momenta around $\pm\frac{\pi}{2}$ where the escape rates are small (and thus the occupation times are large).  The particle typically spends most of the time interval $r\in[0,Nt]$ with momenta $K_{r}\approx \pm \frac{\pi}{2}$  corresponding to small velocities.  When $0\leq\gamma<1 $, then we conjecture  that $N^{-1+\frac{\gamma }{2} }Y_{Nt}$ converges in law to a nontrivial limit.    This would suggest that the quantum model discussed above behaves ballistically when $H^{\prime}(\pm \frac{\pi}{2})\neq 0$, corresponding to the case $\ga=0$.

\subsection{A classical analogue}\label{Sec: classical}

There is a sense in which our present quantum problem differs little
 from an analogous classical problem that starts
from a linear Boltzmann equation.  We explain that analogy here.
Among other things it relates our results to the recent interest
and work on diffusive behavior in systems of coupled oscillators
where energy transport can be understood
as a wave scattered by anharmonicities, \cite{olla}.\\

Consider  a stochastic dynamics with state space $S=
\bbZ^{d}\times \bbT^{d}$ such that the probability density
$P_{t}(x,k)=\Gamma_{t}(P)(x,k)$ evolves as \begin{multline} \label{Classical}
\frac{d}{dt}P_{t}(x,k)= -\sum_{j=1}^{d}\big| (\nabla H)_{j}(k)\big|\big(P_{t}(x,k)-P_{t}(x-s_{j}(k)e_{j},k)\big)\\+\int_{(x',k')\in S}
\big(T(x,k;x',k')P_{t}(x',k')-T(x',k';x,k)P_{t}(x,k)\big), \end{multline}
for an initial $P_{0}(x,k)=P(x,k)$, where the $e_{j}$ are the standard basis vectors of $\bbZ^{d}$, $H:\tor\rightarrow \mathbb{R}$ is differentiable, $s_{j}(k)$ is the sign of $(\nabla H)_{j}(k)$, and    $T(x,k;x',k')\geq 0$
is a transition matrix describing the rates of Poisson timed jumps
from $(x',k')$ to $(x,k)$ and the symbol $\int_{(x',k')\in S}$ stands for $\int_{\tor} \d k' \sum_{x' \in \bbZ^d}$.  The first term on the right generates spatial steps with Poisson rate $|(\nabla H)_{j} (k)|$ in the direction $s_{j}(k)e_{j}$.   That term could be formally absorbed into the second term on the right side of~(\ref{Classical}), although it is analogous to the kinetic term for the quantum case in its mathematical treatment.  If the particle were living on $\bbR^{d}$ rather than $\bbZ^{d}$, then the first term would have the less awkward form $-(\nabla H)(k)\cdot \nabla_{x}P_{t}(x,k)$ for a differentiable $P_{t}(x,k)\in L^{1}(\bbR^{2d})$, which describes a deterministic kinetic motion.  
The connection can be made through a formal limit 
 $$\quad \quad (\nabla H)(k)\cdot \nabla_{x}P_{t}(x,k)=\lim_ {\epsilon\rightarrow 0}     
\frac{1}{\epsilon}\sum_{j=1}^{d}\big| (\nabla H)_{j}(k)\big|\big(P_{t}(x,k)-P_{t}(x-\epsilon s_{j}(k)e_{j},k)\big),\quad x,k\in \bbR^{d},$$
which is essentially a law of large numbers from the stochastic point of view, where many small random jumps (of size $\epsilon\ll 1$) combine to form a deterministic quantity.

  When $T$ satisfies
$T(x+z,k;x'+z,k')=T(x,k;x',k')$, then the master
equation~(\ref{Classical}) describes a formally
translation invariant evolution.
  From a classical physics perspective the evolution \eqref{Classical} is still somewhat strange as it involves
 both jumps in position ($x$) and in momentum ($k$) for the single particle's state evolution.  Still one can
  ask the classical questions about its diffusive behavior.  We must then show that the centered position
  $(x_t - vt)/\sqrt{t}$
  converges in distribution to a Gaussian with non-degenerate covariance matrix. We show here how this can proceed along the very same lines as for the  quantum case.\\

  By translation-invariance the dynamics does not
``feel" the location of the particle and it follows that the
``momentum'' dynamics is Markovian, or, the marginal probability,
$\tilde{P}_{t}(k)=\sum_{x\in \bbZ^{d}}P_{t}(x,k)$, for the $k$
variable obeys an autonomous first order equation: \beq
\label{Marginal} \frac{d}{dt}\tilde{P}_{t}(k)= \int_{ \bbT^{d}
}dk'
\big(\tilde{T}(k,k')\tilde{P}_{t}(k')-\tilde{T}(k',k)\tilde{P}_{t}(k)\big),
\eeq where $\tilde{T}(k,k')=\sum_{z\in\bbZ^{d}} T(x+z,k; x,k') $.   This is
analogous to the zero fiber of our decomposition for the quantum
dynamics. Defining $[P_{t}]_{p}(k)=\sum_{x\in
\bbZ^{d}}e^{\i  px}P_{t}(x,k)$, the dynamics~(\ref{Classical})
operates as $[\Gamma_{t}]_{p}[P]_{p}=[\Gamma_{t}(P)]_{p}$  for some
semigroup $[\Gamma_{t}]_{p}:L^{1}(\bbT^{d})\rightarrow
L^{1}(\bbT^{d})$ whose generator is written below.

The closest thing to a joint probability density for position and momentum in the quantum
case is the Wigner function.  Here we go in the opposite direction and
 we define a ``classical density matrix'' from a joint distribution
 function $P(x,k)$.
That is, for a momentum representation kernel, we formally define:
\beq \label{AntiWigner}
\rho(k_{1},k_{2})= \sum_{x\in\bbZ^{d}}  \,
 e^{ -\i  x(k_{1}-k_{2}) }P\big(x,\frac{k_{1}+k_{2}}{2} \big).
\eeq In this case,  the positivity of $\rho$ is lost, since it is
not in general the kernel of an operator with positive spectrum.
The invariant fibers again correspond to $p=k_{1}-k_{2}$  in the
momentum representation.    The dynamics can thus be written in
its fibers with $\frac{d}{dt}[\Gamma_{t}]_{p}$ operating as
\beq \label{ClassicalFiber} \frac{d}{dt}[P_{t}]_{p}(k)=
-\i h_{p}(k)[P_{t}]_{p}(k) +\int_{\tor} \d k'
\big(\tilde{T}_{p}(k,k')[P_{t}]_{p}(k')-\tilde{T}_{0}(k',k)[P_{t}]_{p}(k)\big)
\eeq
 where $h_{p}(k)=-\i \sum_{j=1}^{d}(1-e^{\i s_{j}(k)p_{j}})|(\nabla H)_{j}(k)|   $ and  $\tilde{T}_{p}(k,k')= \sum_{z\in
\bbZ^{d}}e^{\i   p z}T(x+z,k; x,k')  $.  One notices similarities with
the fiber decomposition and  with \eqref{Fiber: again}.

\section{Proofs}

We need a technical lemma to deal conveniently with trace-class operators
\begin{lemma}\label{lem: acrobatics}
Let $C \in \caB_1(\caH)$, for some Hilbert space $\caH$, and let $X$ be a self-adjoint operator.  If $XC \in \caB_1$ , then
\beq
 \lim_{\ga \searrow 0}\frac{1}{\i\gamma}(e^{\i\gamma X}-I)C   = XC \label{Horse}
\eeq
where the convergence is meant in the sense of $\caB_1(\caH)$.
\end{lemma}
\begin{proof}
Let \beq C=\sum_{n \in \bbN} \lambda_{n}  | f_{n}\rangle\langle g_{n}|, \qquad f_n , g_n \in \caH , \qquad  \sum_n  \str \la_n \str <\infty  \label{Defi}  \eeq be
the singular value decomposition of $C \in \caB_1$. Both families $f_n$ and $g_n$  are an orthonormal set.
By the comment~(\ref{Give}) and $Cg_{n}=\lambda_{n}f_{n}$,  it follows that when $\lambda_{n}\neq 0$, then  $f_n \in \Dom X$.

If there are only finitely many terms in the singular value decomposition~(\ref{Defi}), then the convergence~(\ref{Horse}) is guaranteed by an application of Stone's theorem for each $n$.  When there are an infinite number of terms in~(\ref{Defi}), an extra estimate is required to bound the tail of the sum.

Defining the projection $P_{N}=\sum_{n=1}^{N}|g_{n}\rangle \langle g_{n}|$, then we can write
$$\frac{1}{\i  \gamma}(e^{\i  \gamma X}-I)C- XC= \sum_{n=1}^{N}\lambda_{n}(\frac{1}{\i  \gamma}(e^{\i  \gamma X}-I)- X)|f_n \rangle \langle g_{n}| -X C(I-P_{N})+\frac{1}{\i  \gamma}(e^{\i  \gamma X}-I)C(I-P_{N}). $$
Applying the triangle inequality to the above, we have that
\begin{multline}\label{Turnip}
\|\frac{1}{\i  \gamma}(e^{\i  \gamma X}-I)C- XC\|_{1} \leq \sum_{n=1}^{N}|\lambda_{n}|\| (\frac{1}{\i  \gamma}(e^{\i  \gamma X}-I)- X)|f_n \rangle \langle g_{n}|\|_{1}\\+\|X C(I-P_{N})\|_{1}+\|\frac{1}{\i  \gamma}(e^{\i  \gamma X}-I)C(I-P_{N})\|_{1}.
\end{multline}
Our strategy will be to pick, for each $\epsilon>0$ a  number $N_\ep$ such that the last two terms on the right are bounded by $\ep$, for $N\geq N_\ep$ and arbitrary $\ga$. Since the first term, with $N=N_\ep$ vanishes as $\ga \searrow 0$ by the reasoning above, the claim will follow.

Since $XC \in \caB_1(\caH)$, we have
$\|XC(I-P_{N})\|_{1}\rightarrow 0$, as $N \nearrow \infty$.  We can use this to
bound the third term on the right-hand side of~(\ref{Turnip}).  Note that for
$A=\frac{1}{\i  \gamma}(e^{\i  \gamma X}-I)$ and $B=X$, we have $0\leq
|A|^{2}\leq |B|^{2}$ by functional calculus.  It follows that $ Y^{*} |A|^{2}Y\leq  Y^{*}
|B|^{2} Y $ for any $Y\in \mathcal{B}_{\infty}(\mathcal{H})$, and we will use the case when $Y_{N}=C(1-P_N)$.  Since $\cdot \mapsto \sqrt{\cdot}$
is an operator monotone function, we have $ \sqrt{Y^{*}_{N} |A|^{2}
Y_{N}}  \leq \sqrt{Y_{N}^{*} |B|^{2} Y_{N}} $.  With this equality and two applications of the definition of the trace norm, we have
$$\|AY_{N}\|_{1}= \Tr \sqrt{ Y^{*}_{N} |A|^{2} Y_{N}} \leq \Tr \sqrt{Y^{*}_{N} |B|^{2} Y_{N}}= \|BY_{N}\|_{1}.$$
It follows that $\|AY_{N}\|_{1}\rightarrow 0$ as $N\rightarrow \infty$ which completes the proof.

\end{proof}

We continue with the proof of Proposition \ref{lem: fiber}\\

\noindent[Proof of Proposition \ref{lem: fiber}.]\\
\noindent\emph{Step 1} \\
Assume the singular value decomposition for $C$, as in \eqref{Defi}. Then
\beq \label{def: upsilon}
k \mapsto \Upsilon_p(k):= \sum_{n \in \bbN} \lambda_{n}  (\e^{\i \frac{p}{2} X} f_n)(k)   \overline{(\e^{-\i \frac{p}{2} X} g_n) (k)  }
\eeq
is a $L^1$-function (by Cauchy-Schwartz, since it is summable series of products $L^2$-functions), which depends continuously on $p$, since $p \mapsto \e^{\i \frac{p}{2} X}$ is a strongly continuous group on $L^2(\tor)$.
It is straightforward to verify that $ [C]_p := \Upsilon_p $ satisfies our definition of the fiber decomposition \eqref{Fiber: first}. In other words,
\beq
C(k_1,k_2):= [C]_{k_2-k_1} (\frac{k_1+k_2}{2})
\eeq
is a kernel for the operator $C$.  \\

\noindent\emph{Step 2}\\
We show first that, under assumption \eqref{eq: ass moments exist}, $[C]_p$ is actually in $\caC_1$.
First, we show that, If both $C$ and $XC, CX$ are in $\caB_1$, then
\beq \label{eq: how to differentiate traceclass}
\frac{\i  }{2}[\{X_j,C\}]_{p}=\frac{\partial}{\partial p_j}[C]_{p}.
\eeq
By Step 1, It suffices to show that, in $\caB_1$,
\beq \label{eq: onesided conv}
 \frac{1}{\i  \ga} (\e^{\i \ga X}-1)  C  \mathop{\longrightarrow}\limits_{\ga \downarrow 0}     XC
\eeq
which is proven in Lemma \ref{lem: acrobatics}.
The continuity of  $\frac{\partial}{\partial p_{j}}[C]_{p}$  in $p$, follows from \eqref{eq: how to differentiate traceclass} and the general argument in \emph{Step 1}, with $C$ replaced by $\frac{1}{2} \{C,X \}$.  \\

\noindent\emph{Step 3}\\The existence and continuity of the second derivative follows by repeating \emph{Step 2}, since
$  \{ C, X\} X$ and $ X \{ C, X\} $ are in $\caB_1$.

\qed

The following lemma lays out the standard perturbation
theory~\cite{Kato} that we make use of.
\begin{lemma}\label{lem: pert}
Consider a family of bounded operators $L_{p}$ on a Banach space for
$p\in\bbT^{d}$.    Assume that the $\frac{\partial^2}{\partial
p_{i}\partial p_{j}}L_p$ are bounded and continuous as a functions
of $p$ in some neighborhood of $0 \in \bbT^d$.   Finally, assume that $\mathrm{spec} (L_{0})$, the spectrum of $L_0$, contains $0$ as an isolated point, corresponding to a  simple eigenvalue


Then, for $\str p \str$ small enough,  the operator $L_p$ has a unique
eigenvalue $D_p$  such that \beq \label{perturbation series} D_p
=   ( p,  \mathrm{tr}  [ T^{(1)}  P_0  ] )        +    ( p,   \mathrm{tr}  [
T^{(2)}  P_0  ] p )   -   ( p,    \mathrm{tr}  [  T^{(1)}  S
T^{(1)}  P_0  ] p )+  o(p^2), \eeq where $T^{(1)}$ and
$T^{(2)}$ are, respectively, a vector of operators and a self-transpose matrix of operators (i.e. $T^{(2)}_{i,j}=T^{(2)}_{j,i}$), defined by the expansion \beq
L_p-L_0 = (p, T^{(1)}) +
( p,  T^{(2)}p)    +o(p^2), \nonumber \eeq $P_{0}$ is the projection
corresponding to the eigenvalue   $0$  of $L_{0}$, and $S$ is the
reduced resolvent of $L_{0}$, at the eigenvalue $0$, i.e.\ the
solution of
$$
S (0-L_0) =    (0-L_0) S = 1-P_0 , \qquad  S P_0=P_0 S =0.
$$
Moreover, $\mathrm{spec} (L_p) \setminus \{D_p\}$ lies at a distance $o(p^0)$ from $\mathrm{spec} (L_0) \setminus \{ 0\}$.
\end{lemma}

For our model the projection $P_{0}$ has the
form $P_{0}=|\mathcal{P}\rangle \langle 1_{\tor} |$, where $\mathcal{P}$ is
the stationary density for the Markov process of the zero fiber and $1_{\tor}$ is the indicator function over $\bbT^{d}$.   The first-order perturbation term $T^{(1)}$ is given in \eqref{FirstOrder}.
 The second-order perturbation $T^{(2)}$ is given by \eqref{t2} and it only depends on the map $\Psi$, not on $H$.
With the assumption of~(\ref{Symmetry}) for $W=UV$,  so that the second term
on the right-hand side of~(\ref{FirstOrder}) is zero, we have the
following explicit expressions
\begin{eqnarray}
\mathrm{tr}  [ T^{(1)}  P_0  ]&=& \i \int_{\bbT^{d}}dk\, \mathcal{P}(k) (\nabla H)(k)  ,\\
\mathrm{tr}  [ T^{(2)}  P_0 ] & =&- \int_{\bbT^{d}}\,dk\,
\int_{\bbT^{d}}\,
d\nu(\theta) \, \mathcal{P}(k) m_{\theta}^{(2)}(k), \label{sec}  \\
 \mathrm{tr}   [ T^{(1)}  S  T^{(1)}  P_0  ]
 &=&
 -\int_{0}^{\infty}dt
 \int_{\tor}  d k  \, \zeta(k)
  (e^{tL_{0}}\zeta\mathcal{P})(k) \label{third},
 \end{eqnarray}
with the function $\zeta$ as defined in Proposition \ref{pos}. We have used that on the range of $S$, the integral
$\int_{0}^{\infty}dt e^{tL_{0}}$ is well-defined and equal to $S$.

The expression~(\ref{sec}) is non-positive since the matrices $\nabla_{1}\nabla_{2}M_{\theta}(k,k),$ are non-negative, as explained in Section \ref{sec: symmetries}. Seeing that the
expression~(\ref{third}) is non-positive is a little more tricky,
but it helps to rewrite the right-hand side as the following:
$$\int_{0}^{\infty}dt \int_{\tor} dk\,(e^{tL^{*}_{0}} \zeta)(k)
\  \zeta(k)\mathcal{P}(k)=\int_{0}^{\infty}dt\, \bbE_{\mathcal{P}}
[\zeta(Y_{t})  \zeta(Y_{0})].   $$ The evolution has
been shifted to operate on the observables which can then be
rewritten in terms of the expectation $\bbE_{\mathcal{P}} $ of the
Markov process $Y_{t}$ started from the stationary
distribution $\caP$.   We show the non-negativity of this expression in the proof of Proposition~\ref{pos}.\\

\noindent[Proof of Theorem \ref{Main}.]\\

To prove convergence in distribution for $\mu_{t}$, we show
pointwise convergence of the characteristic functions. \beq
\varphi_{\mu_{t}}(\gamma):=\int_{\bbR^d} \d\mu_{t}(x)e^{\i x\gamma}=  \Tr[
\rho_t     e^{\i \frac{\ga}{\sqrt{t} } (X -v t)}]
=e^{ -\i  \sqrt{t}v\gamma} \langle 1_{\tor},
[e^{tL}(\rho)]_{\frac{\gamma}{\sqrt{t}}} \rangle  \label{def:
characteristic as product}\eeq where the third equality makes use
of the fiber decomposition of our dynamics.  In particular, we
used the relation (cf. the proof of Proposition \ref{lem: fiber}) \beq
\label{eq: shift of fiber}  \left[\e^{\i \frac{\ga}{2} X} C \e^{\i
\frac{\ga}{2} X} \right]_p=   \left[ C \right]_{p+\ga}, \qquad   C
\in \caB_1. \eeq

   For a fixed $\gamma$  the limit involves only
small fibers $p\propto t^{-\frac{1}{2}}$, which suggests using a
perturbation argument around the zero fiber.  The continuity of the second derivatives of $L_{p}$ follows from the continuity assumptions on the second derivatives of the functions $M_{\theta}$ and $H$ in Assumption~\ref{as1} and the forms~(\ref{Fiber: again}) and~(\ref{Fiber: again2}).   Lemma~\ref{lem: pert}, for a small enough neighborhood of $p=0$,
say $\caU \subset \tor$, we have that \beq \label{eq: exp
convergence} [\e^{t L}\rho]_p=\e^{t L_{p} }[\rho]_p   = P_p \e^{t D_p } +  (1-P_p) \e^{t
V_p} [\rho]_p, \qquad   V_p := L_p -D_p P_p \eeq and $\norm \e^{t
V_p} \norm =O(\e^{-t b})$ as  $t \nearrow \infty$, for some $b>0$
satisfying $b-b_A \rightarrow 0$ as $p \searrow 0$. The norm
refers to the operator norm on $\caB(L^1(\tor))$ and $b_A$ is the
gap of the operator $A$ (Assumption \ref{ass: irreducibility}).

We now show that, in $L^1(\tor)$,
\beq \label{eq: conv in lone}
 e^{ -\i  \sqrt{t}v\gamma} [\e^{t L}\rho]_{\frac{\gamma}{\sqrt{t}}}
 \quad    \mathop{\longrightarrow}\limits_{t \nearrow \infty}
  \quad   \e^{- \frac{1}{2}(\ga, \si \ga)}  \caP
\eeq
This follows by combining \eqref{eq: exp convergence}, the relation
\beq \label{expansion dp} D_p =  \i (p,v) - \frac{1}{2}(p, \sigma p)+o(p^2) \eeq
 (which follows from Lemma \ref{lem: pert}),  and the fact that
\beq \label{two lone convergence} P_p \quad
\mathop{\longrightarrow}\limits_{t \nearrow \infty}  \quad P_0=
\str   \caP \rangle \langle 1_{\tor}\str , \qquad   [\rho]_p
\quad \mathop{\longrightarrow}\limits_{t \nearrow \infty}  \quad
[\rho]_0 \eeq where the first convergence is  in $\caB(L^1(\tor))$
and the second in $L^1(\tor)$. The first  claim of \eqref{two lone
convergence} follows again from Lemma \ref{lem: pert}, the second
is a consequence of Proposition \ref{lem: fiber}. We have shown
\eqref{eq: conv in lone}.  Since norm convergence implies weak
convergence, in particular
$[e^{tL}(\rho)]_{\frac{\gamma}{\sqrt{t}}}$ integrated against the
indicator $1_{\bbT^d}$ converges to the desired value.     Hence
$\mu_{t}$ converges in distribution.

We now prove the convergence of the first and second
moments~(\ref{ginger}).  By \eqref{def: characteristic as
product}, and the usual connection between moments and derivatives
of the characteristic function, we have (whenever the
right-hand side exists),
\begin{multline}\label{Coco}
\frac{1}{t}\sum_{x \in \lat}  \rho_{t}(x,x)(x_i-tv_i)(x_j-tv_j)
\\=\frac{1}{t}\big(-\frac{\partial^{2}}{\partial p_{i}\partial
p_{j}}\langle 1, \e^{t
L_{p}}[\rho_{0}]_{p}\rangle\Big|_{p=0}+\frac{\partial}{\partial
p_{i}}\langle 1, \e^{t L_{p}}[\rho_{0}]_{p}\rangle\Big|_{p=0}
\frac{\partial}{\partial p_{j}}\langle 1, e^{t
L_{p}}[\rho_{0}]_{p}\rangle\Big|_{p=0}\big)
\end{multline}
 Note that since the operator $L_{p}$ has two continuous derivatives, the operators $P_p$,$V_p$ (defined above) and the eigenvalue $D_p$ do also.  In particular $\norm ( \frac{\partial^{2} }{\partial p_{i}\partial p_{j} })  V_p \norm$ is bounded for $p \in \caU$.
One can see that, \[
 \sup_{p \in
  \caU} \norm ( \frac{\partial^{2} }{\partial p_{i}\partial p_{j}})
   \e^{t V_p}  \norm =  O( t^2 \e^{-b t})  ,\quad   t \nearrow \infty.
\] (and  a similar bound for the first-order derivatives).  By
Lemma \ref{lem: fiber}, the function $p \mapsto [\rho]_p$ is
$\caC^2$ and we obtain \[
  \frac{\partial^{2} }{\partial p_{i}\partial p_{j} }
   \big(\e^{t V_p} [\rho]_p\big)      \Big\str_{p \in \caU}  \quad
       \mathop{\longrightarrow}\limits_{t \nearrow \infty}   \quad 0.
\]
 Hence,  in~(\ref{Coco}) we can replace $e^{t L_{p}}$ with
$P_{p}\e^{tD_{p}}$, for large $t$, and we see the convergence to
$\sigma$ using the expansion \eqref{expansion dp}.

\qed

 \noindent[Proof of Proposition \ref{pos}.] \\

\noindent \emph{Proof of Statement 1) }
In the proof of Theorem \ref{Main}, we showed the pointwise convergence of the characteristic function $\varphi_{\mu_{t}}$, i.e.,
\beq
\varphi_{\mu_{t}}(\gamma)=\int_{\bbR^d} \d\mu_{t}(x)e^{\i x\gamma}  \qquad \mathop{\longrightarrow}\limits_{t \uparrow \infty}  \qquad \e^{-\frac{1}{2}( \gamma ,\sigma \gamma)}
\eeq
(see e.g.\  \eqref{def: characteristic as product} and \eqref{eq: conv in lone}). 
Suppose there were a $\gamma \in \bbR^d$ such
that $(\gamma, \sigma \gamma) <0$, then, for large
enough $t$,  $\varphi_{\mu_{t}}(\gamma) >1$, 
which is impossible since $\varphi_{\mu_{t}}$ is the characteristic function of a probability measure.
This proves the non-negativity of the diffusion
matrix $\sigma$.
\qed
\vspace{5mm} \\
\emph{Proof of Statement 2) } Now we consider the non-negativity of the matrix $\alpha$.  We show that $\alpha$ is a non-negative matrix by showing that the expression $( w, \alpha w)$ for $w \in \bbR^d$ is always non-negative.  Using an unsymmetrized form for $\alpha$, we can rewrite our evaluation as
\begin{align}\label{Evaluated}
( w, \alpha w)=\int_{0}^{\infty}\d t\,
\bbE_{\mathcal{P}}[f(Y_{t})f(Y_{0})],
\end{align}
 where  $f(k):=( w, \zeta(k) ) $ is real valued.   Let the stationary Markov process $Y_{t}$ be extended to all negative values of $t$.  Then $G(t)=\bbE_{\mathcal{P}}[f(Y_{t})f(Y_{0})]$ is an even function and so~(\ref{Evaluated}) is twice the value of the Fourier transform $\tilde{G}(z)$ of $G(t)$ at $z=0$.   We will show that $\tilde{G}(z)$ is non-negative valued.   Using Bochner's theorem we just need to check that $G(t-s)$ defines
a positive operator on $L^{2}(\mathbb{R})$.   Let $\eta\in
L^{2}(\mathbb{R})$, then
\[
\int_{\mathbb{R}^{2}}dt\,ds\,
 \bar{\eta}(t)G(t-s)\eta(s)= \bbE_{\caP}[ |\int_{\mathbb{R}} dt\, \eta(t) f(Y_{t})|^{2} ]
 \]
where we have used the stationarity property
$\bbE_{P}[f(Y_{t-s})f(Y_{0})] = \bbE_{\caP}[f(Y_{t})f(X_{s})]$.
\qed
\vspace{5mm} \\
\emph{Proof of Statement 3) } The strict positivity of $\al$ is established as follows. Let $w \in \mathbb{R}^d$. By the assumption that the velocity fluctuates, $g(k):=    ( w, \zeta(k) )   \in \caH_\caP$ satisfies
\[
g \neq 0 , \qquad    \langle 1_{\tor},  g \rangle_\caP =0 \] where
$1_{\tor} \in \caH_\caP$, the identity function on $\tor$, is the
$0$-eigenvector of $A_\caP$.  Note further that \beq ( w, \alpha
w) =    \langle g , (A_\caP)^{-1} g \rangle_\caP, \eeq since one
can easily check that $h:= (A_\caP)^{-1} g  \in \caH_\caP$ by
using the fact that $A_\caP$ is bounded and $A$ has a gap.
  Assume that $\al$ is not strictly positive. Then, there is a $w \in \mathbb{R}$ such that (recall that $g$ and hence $h$ depend on $w$)
\[    \langle g , (A_\caP)^{-1} g \rangle_\caP=        \langle A_\caP h,  h \rangle_\caP=0. \]
In particular, this implies that
\beq \label{eq: real part zero} \langle   \mathrm{Re} (A_\caP) h,  h \rangle_\caP= 0. \eeq
Since $-\mathrm{Re} (A_\caP) $ is a positive operator (as follows from the fact that $A_\caP$ is a Markov generator), we get  $\mathrm{Re} (A_\caP) h=0$.
By the sectoriality assumption, it follows that also  $\mathrm{Im} (A_\caP) h=0$ and hence $A_\caP h=0$. Indeed,  for $\la>0$ and $v \in \caH_p$,
\[
 \str   \langle (h+\la v) , \mathrm{Im} (A_\caP)  (h+\la v) \rangle_\caP  \str  \leq \ga    \langle (h+\la v) , \mathrm{Re} (A_\caP)  (h+\la v) \rangle_\caP  =  \ga\la^2 \langle v , \mathrm{Re} (A_\caP)  v \rangle_\caP
\]
and hence the  $O(\la)$-term  has to vanish on the left hand side
for all $v$. Since the zero-eigenvector of $A_\caP$ is unique by
assumption, we obtain $h =c  1_{\tor}, c \in \bbR$.  This leads to
a contradiction with the fact that $h=(A_\caP)^{-1}g$ and $\langle
1_{\tor}, \caP \rangle=0$. Hence $\al$ is strictly positive. \qed
\vspace{5mm}\\

Note that \eqref{Evaluated} is related to the familiar central
limit theorem for Markov processes
\[
\lim_{T\rightarrow \infty}\frac{1}{\sqrt{T}}\int_{0}^{T}dt\,
f(Y_{t}) \rightarrow \mathcal{N}(0,\sigma) \]
where the convergence
is in distribution, $\sigma=\int_{0}^{\infty}dt\,
\bbE_{\mathcal{P}}[f(Y_{t})f(Y_{0})],$ and the function $f$
satisfies $\int_{\bbT^{d}} dk\,f(k)\mathcal{P}(k)dk=0$.

\section*{Acknowledgement}

At the time when this work was completed, W.D.R was a postdoctoral fellow supported by the FWO-Flanders.  J.C. has benefited by support from the Belgian Interuniversity Attraction Pole P6/02 and the European Research Council grant No. 227772.

 \end{document}